# OpenPh - Numerical Physics Library

Authors: George Milescu [1], Gabriel Noaje [1], Florin Pop [2]
*Politehnica University, Bucharest*

## Abstract

**Keywords**: numerical physics, MATLAB, quantum physics, classical mechanics, module, photoelectric effect, Schrödinger equation, vibration of a string

Numerical physics has gained a lot of importance in the last decade, its efficiency being motivated and sustained by the growth of computational power. This paper presents a concept that is to be developed in the next few years: OpenPh. OpenPh is a numerical physics library that makes use of the advantages of both open source software and MATLAB programming. Its aim is to deliver the instruments for providing numerical and graphical solutions for various physics problems. It has a modular structure, allowing the user to add new modules to the existing ones and to create its own modules according to its needs, being virtually unlimited extendable. The modules of OpenPh are implemented using MATLAB engine because it is the best solution used in engineering and science, providing a wide range of optimized methods to accomplish even the toughest jobs.

Current version of OpenPh includes two modules, the first one providing tools for quantum physics and the second one for mechanics. The quantum physics module deals with the photoelectric effect, the radioactive decay of carbon-11, and the Schrödinger equation - particle in a box. The classical mechanics module includes the study of the uniform circular motion, the forced damped harmonic oscillations and the vibration of a fixed-fixed string.

**Introduction**

The foundation for this physics library is the well-know computational engine Matlab. *Why Matlab?* Because it is the award-winning solution used in engineering and science providing a wide range of means to accomplish even the toughest jobs. It modularity and facile expansion allows every user to develop its own modules and application to fulfill its needs.

The library presented is far from being complete, but due to its modularity and scalability it allows virtually unlimited extension. Currently it includes functions for solving problems from two different domains of the physics: mechanics and quantum physics. In addition, it includes some general functions basically oriented on mathematics. These functions can be used by any of the specialized modules in order to simplify calculations or as stand-alone modules.

*And why OpenPh?* OpenPh comes from Open Physics and states the idea that lead to the development of this library: an open-source freely distributable modular library that uses Matlab in physics domain.

The two modules currently available in the library deliver the following specific functions:

A. *Quantum physics:* photoelectric effect, simulated radioactive decay of carbon-11, Schrödinger equation - particle in a box
B. *Mechanics:* uniform circular motion, forced damped harmonic oscillations, simple harmonic oscillator (pendulum), vibration of a fixed-fixed string
C. *Relativity* – under development

**A. Quantum physics**

---

[1] Automatic Control and Computers Faculty
[2] Engineer

Quantum mechanics is a fundamental physical theory which extends and corrects Newtonian mechanics, especially at the atomic and subatomic levels. The library offers tools to analyze and calculate some common problems in quantum theory.

## A1. Photoelectric effect

The photoelectric effect is the emission of electrons from a surface (usually metallic) upon exposure to, and absorption of, electromagnetic radiation (such as visible light and ultraviolet radiation) that is above the threshold frequency particular to each type of surface. No electrons are emitted for radiation below the threshold frequency, as they cannot gain sufficient energy to overcome their atomic bonding.

In analyzing the photoelectric effect quantitatively using Einstein's method, the following equivalent equations are used:

$$hf = hf_0 + \frac{1}{2}mv_m^2$$

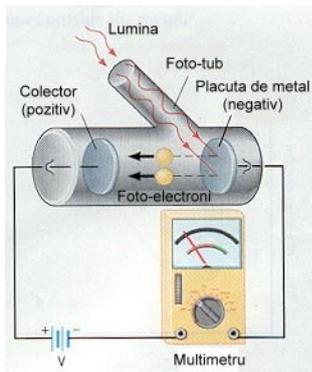

Figure 1 – Experiment for the photoelectric effect

where $hf$ is the energy of photon, $hf_0$ is the energy needed to remove an electron and $\frac{1}{2}mv_m^2$ is the kinetic energy of the emitted electron.

In order to present the photoelectric effect the circuit in figure 1 is realized. The library function calculates the voltage when no current is passing through the multimeter.

## A2. Simulated radioactive decay of carbon-11

Radioactive decay is the set of various processes by which unstable atomic nuclei (nuclides) emit subatomic particles.

The decay of an unstable nucleus (radionuclide) is entirely random and it is impossible to predict when a particular atom will decay. However, it is equally likely to decay at any time. Therefore, given a sample of a particular radioisotope, the number of decay events expected to occur in a small interval of time *dt* is proportional to the number of atoms present. If *N* is the number of atoms, the following first-order differential equation can be written:

$$\frac{dN}{dt} = -\lambda N$$

Particular radionuclides decay at different rates, each having its own decay constant (λ). The negative sign indicates that N decreases with each decay event. The solution to this equation is the following function:

$$N(t) = N_0 e^{-\lambda t}$$

This function represents exponential decay. It is only an approximate solution, for two reasons. Firstly, the exponential function is continuous, but the physical quantity N can only take positive integer values. Secondly, because it describes a random process, it is only statistically true. However, in most common cases, N is a very large number and the function is a good approximation.

The library function tries to express this approximation presenting the theoretical function and the randomly generated number of decayed nuclei for Carbon 11 atom. (figure 2,3)

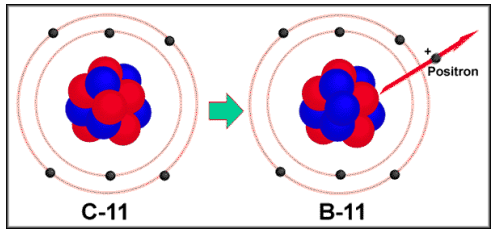 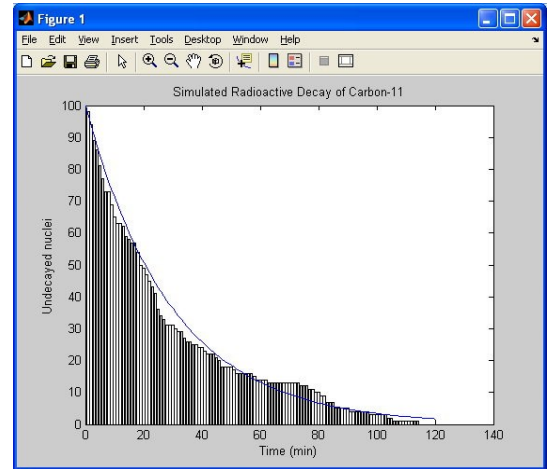

Figure 2 – Radioactive decay of carbon-11 (scheme)

Figure 3 - Simulated radioactive decay of carbon-11

### A3. Schrödinger equation - particle in a box

In physics, the particle in a box (or the square well) is a simple idealized system that can be completely solved within quantum mechanics. It is the situation of a particle confined within a finite region of space (the box) by an infinite potential that exists at the walls of the box. The particle experiences no forces while inside the box, but is constrained by the walls to remain in the box.

The quantum behavior in the box includes:

1. **Energy quantization** - It is not possible for the particle to have any arbitrary definite energy. Instead only discrete definite energy levels are allowed (if the state is not a steady state, however, any energy past zero-point energy is allowed on average).
2. **Zero-point energy** - The lowest possible energy level of the particle, called the zero-point energy, is nonzero.
3. **Nodes** - In contrast to classical mechanics, the Schrödinger equation predicts that for some energy levels there are nodes, implying positions at which the particle can never be found.

The Schrödinger equation is:

$$H(t)|\Psi(t)\rangle = i\hbar \frac{\partial}{\partial t}|\Psi(t)\rangle$$

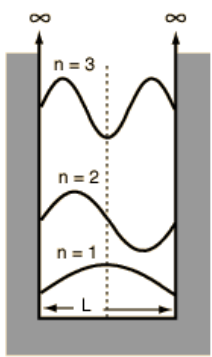

Analytical solutions of the time-independent Schrödinger equation can be obtained for a variety of relatively simple conditions. One of these simple conditions is the particle in a 1-dimensional box. (figure 4)

The library function gives the solution to the equation for 4 different types of boxes (square well, double well, parabolic, free-hand). (figure 5)

Figure 4 – Particle in a box with infinite walls

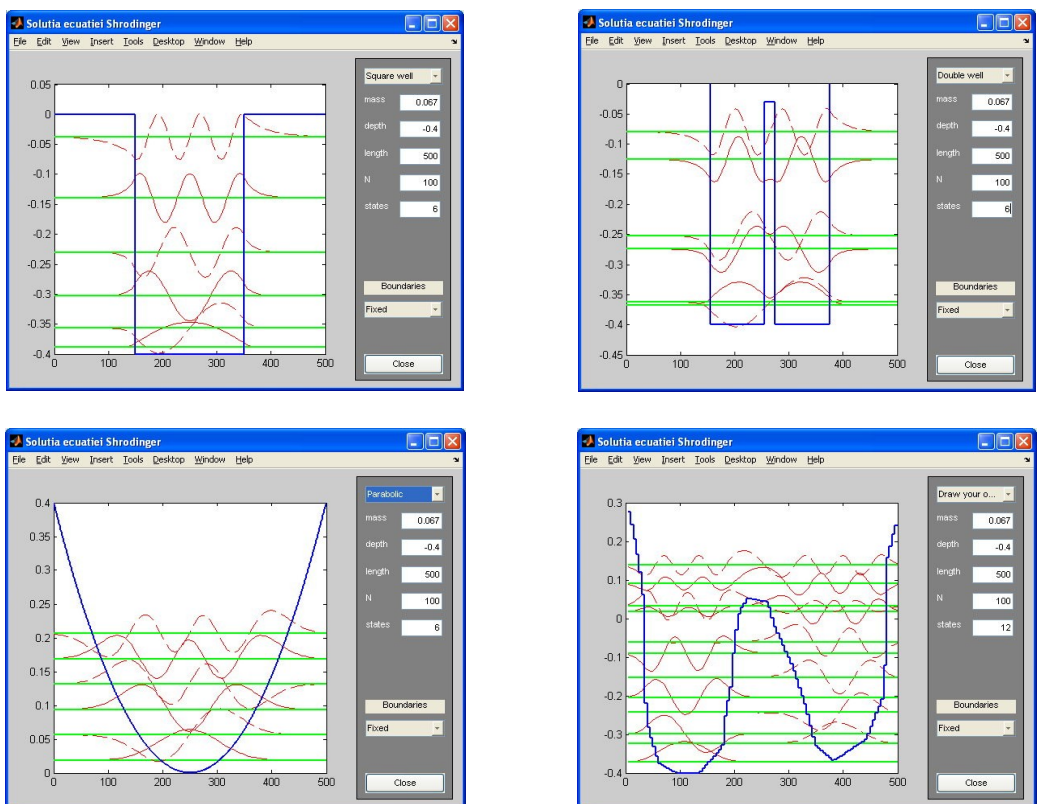

Figure 5 - Particle in a 1-dimensional box

## B. Classical mechanics

Classical mechanics is one of the two major sub-fields of study in the science of mechanics, which is concerned with the motions of bodies, and the forces that cause them. The library offers tools to analyze and calculate some common problems in classical mechanics.

### B1. Uniform circular motion

Circular motion is movement with constant speed around a circle. (figure 6). Circular motion can be described by means of parametric equations:

$$x(t) = R \cos \omega t$$
$$y(t) = R \sin \omega t$$

A function for circular motion was included in the library in order to represent an intuitive path. (figure 7)

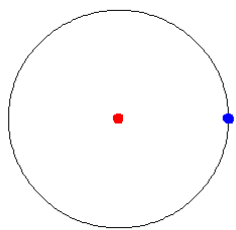 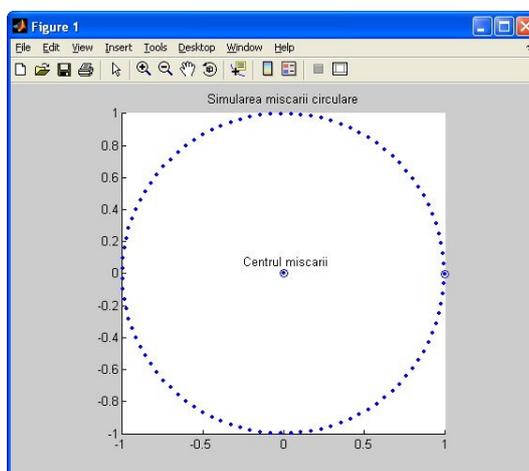

Figure 6 – Circular motion path    Figure 7 – Uniform circular motion simulation

*B2. Forced damped harmonic oscillations*

Forced damped harmonic oscillator satisfies equation:

$$m\frac{d^2x}{dt^2} + r\frac{dx}{dt} + kx = F_0 \cos(\omega t)$$

Derived from this general equation simpler oscillators exist. The library offers a set of powerful functions allowing simulating and calculating different elements of an oscillator such as amplitude, angular frequency and acceleration. Each of these elements can be graphically represented. Figure 8 shows such an example.

Furthermore, there are used 2 different ways of computing the elements of the oscillator, that's why a special function has been implemented to make 4 plots comparing the small amplitude solution with the numerical solution. (figure 9)

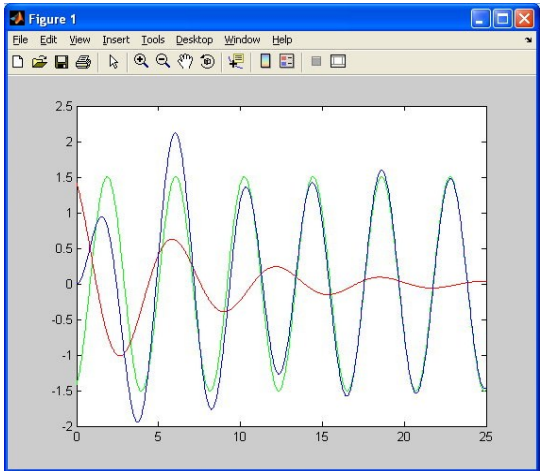
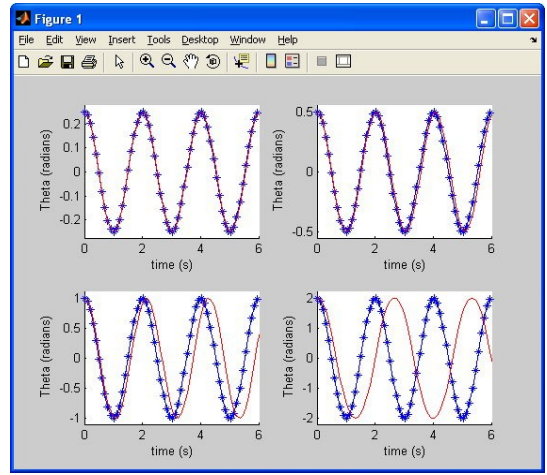

Figure 8 – Amplitude, angular frequency, acceleration of an oscillator

Figure 9 – Comparison graphs

*B3. Vibration of a fixed-fixed string*

When the end of a string is fixed, the displacement of the string at that end must be zero. A transverse wave traveling along the string towards a fixed end will be reflected in the opposite direction. When a string is fixed at both ends, two waves traveling in opposite directions simply bounce back and forth between the ends.

$$y(x,t) = y_m \sin(kx - \omega t) + y_m \sin(kx + \omega t)$$

A string which is fixed at both ends will exhibit strong vibrational response only at the resonance frequencies

$$f_n = \frac{nv}{2L}$$

where $v = \sqrt{\frac{\tau}{\rho}}$ is the speed of transverse mechanical waves on the string, *L* is the string length, and *n* is an integer. At any other frequencies, the string will not vibrate with any significant amplitude. The resonance frequencies of the fixed-fixed string are harmonics (integer multiples) of the fundamental frequency (n=1).

The vibrational pattern (mode shape) of the string at resonance will have the form:

$$y(x,t) = y_m \sin\left(\frac{n\pi}{L}\chi\right)\cos(2\pi f t)$$

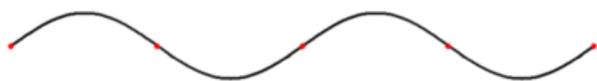

Figure 10 - Standing wave in stationary medium (the red dots represent the wave nodes)

This equation represents a standing wave. There will be locations on the string which undergo maximum displacement (antinodes) and locations which to not move at all (nodes). In fact, the string may be touched at a node without altering the string vibration. (figure 10)

The library has a function that shows the motion of such string in a very realistic manner. (figure 11)

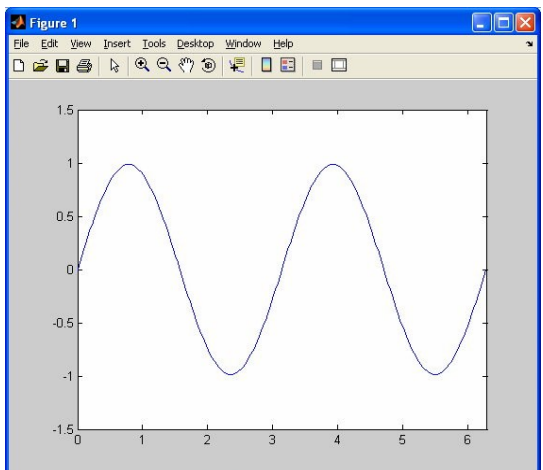

Figure 11 – String vibration simulation

Beside the specialized modules there are also some generic functions such as:

- function that generates Fahrenheit-Celsius conversion table
- function that generates comparison table between factorial and Stirling approximation

**Conclusions**

Although the library is not complete it offers a wide range of tools for solving different types of problems. It has a user-friendly interface and it is extremely easy to use.

Relying on the powerful Matlab engine, functions can work with large sets of data.

Functions have practical applications.

Modularity allows trouble free expansion, add of new modules or optimization of the old ones.